\documentclass[11pt,a4paper]{article}
\usepackage[hyperref]{acl2019}
\usepackage{times}
\usepackage{latexsym}
\usepackage{graphicx}
\usepackage{hyperref}
\usepackage{booktabs}
\usepackage{url}

\usepackage{url}

\aclfinalcopy 

\title{Learning Open Information Extraction of Implicit Relations from Reading Comprehension Datasets}

\author{Jacob Beckerman \\
  The Wharton School \\
  University of Pennsylvania \\
  \texttt{jbecke@wharton.upenn.edu} \\\And
  Theodore Christakis \\
  Western Engineering \\
  University of Western Ontario \\
  \texttt{tchris22@uwo.ca} \\}

\date{}

\begin{document}
\maketitle
\begin{abstract}

The relationship between two entities in a sentence is often implied by word order and common sense, rather than an explicit predicate. For example, it is evident that ``Fed chair Powell indicates rate hike'' implies (Powell, is a, Fed chair) and (Powell, works for, Fed). These tuples are just as significant as the explicit-predicate tuple (Powell, indicates, rate hike), but have much lower recall under traditional Open Information Extraction (OpenIE) systems. Implicit tuples are our term for this type of extraction where the relation is not present in the input sentence. There is very little OpenIE training data available relative to other NLP tasks and none focused on implicit relations. We develop an open source, parse-based tool for converting large reading comprehension datasets to OpenIE datasets and release a dataset 35x larger than previously available by sentence count. A baseline neural model trained on this data outperforms previous methods on the implicit extraction task.

\end{abstract}

\section{Introduction}
Open Information Extraction (OpenIE) is the NLP task of generating (subject, relation, object) tuples from unstructured text e.g. ``Fed chair Powell indicates rate hike'' outputs (Powell, indicates, rate hike). The modifier \emph{open} is used to contrast IE research in which the relation belongs to a fixed set. OpenIE has been shown to be useful for several downstream applications such as knowledge base construction \cite{okr}, textual entailment \cite{entailment}, and other natural language understanding tasks \cite{IEintermediate}. In our previous example an extraction was missing: (Powell, works for, Fed). Implicit extractions are our term for this type of tuple where the relation (``works for'' in this example) is not contained in the input sentence. In both colloquial and formal language, many relations are evident without being explicitly stated. However, despite their pervasiveness, there has not been prior work targeted at implicit predicates in the general case. Implicit information extractors for some specific implicit relations  such as noun-mediated relations, numerical relations, and others \cite{Demonyms, Bootstrapping, conjunctiveIE} have been researched. While specific extractors are important, there are a multiplicity of implicit relation types and it would be intractable to categorize and design extractors for each one. 

Past general OpenIE systems have been plagued by low recall on implicit relations \cite{SupervisedOIE}. In OpenIE's original application --  web-scale knowledge base construction -- this low recall is tolerable because facts are often restated in many ways \cite{origionalOpenIE}. However, in downstream NLU applications an implied relationship may be significant and only stated once \cite{IEintermediate}. 

The contribution of this work is twofold. In Section 4, we introduce our parse-based conversion tool and convert two large reading comprehension datasets into implicit OpenIE datasets. In Section 5 and 6, we train a simple neural model on this data and compare to previous systems on precision-recall curves using a new gold test set for implicit tuples.

\section{Problem Statement}

We suggest that OpenIE research focus on producing implicit relations where the predicate is not contained in the input span. Formally, we define implicit tuples as (subject, relation, object) tuples that:

\begin{enumerate}
  \item Have a subject and object word or phrase contained in the input sentence.
  \item Have a relation token(s) entailed by word order of the sentence but not contained in it.
\end{enumerate}

These ``implicit'' or ``common sense'' tuples reproduce the relation explicitly, which may be important for downstream NLU applications using OpenIE as an intermediate schema. For example, in Figure 1, the input sentence tells us that the Norsemen swore fealty to Charles III under ``their leader Rollo''. From this our model outputs (The Norse leader, was, Rollo) despite the relation never being contained in the input sentence. Our definition of implicit tuples corresponds to the ``frequently occurring recall errors'' identified in previous OpenIE systems \cite{SupervisedOIE}: noun-mediated, sentence-level inference, long sentence, nominalization, noisy informal, and PP-attachment. We use the term 
\textit{implicit tuple} to collectively refer to all of these situations where the predicate is absent or very obfuscated.

\section{Related Work}

\subsection{Traditional Methods}

Due to space constraints, see Niklaus et al. \shortcite{Survey} for a survey of of non-neural methods. Of these, several works have focused on pattern-based implicit information extractors for noun-mediated relations, numerical relations, and others \cite{Demonyms, Bootstrapping, conjunctiveIE}. In this work we compare to OpenIE-4 \footnote{https://github.com/knowitall/openie}, ClausIE \cite{clausie}, ReVerb \cite{ReVerb2011}, OLLIE \cite{ollie}, Stanford OpenIE \cite{stanford}, and PropS \cite{propS}.

\subsection{Neural Network Methods}

Stanovsky et al. \shortcite{SupervisedOIE} frame OpenIE as a BIO-tagging problem and train an LSTM to tag an input sentence. Tuples can be derived from the tagger, input, and BIO CFG parser. This method outperforms traditional systems, though the tagging scheme inherently constrains the relations to be part of the input sentence, prohibiting implicit relation extraction. Cui et al. \shortcite{NeuralOpenIE} bootstrap (sentence, tuple) pairs from OpenIE-4  and train a standard seq2seq with attention model using OpenNMT-py \cite{OpenNMT}. The system is inhibited by its synthetic training data which is bootstrapped from a rule-based system.

\subsection{Dataset Conversion Methods}

Due to the lack of large datasets for OpenIE, previous works have focused on generating datasets from other tasks. These have included QA-SRL datasets \cite{benchmark} and QAMR datasets \cite{SupervisedOIE}. These methods are limited by the size of the source training data which are an order of magnitude smaller than existing reading comprehension datasets.

\section{Dataset Conversion Method}

Span-based Question-Answer datasets are a type of reading comprehension dataset where each entry consists of a short passage, a question about the passage, and an answer contained in the passage. The datasets used in this work are the Stanford Question Answering Dataset (SQuADv1.1) \cite{Squad} and NewsQA \cite{newsQA}. These QA datasets were built to require reasoning beyond simple pattern-recognition, which is exactly what we desire for implicit OpenIE. Our goal is to convert the QA schema to OpenIE, as was successfully done for NLI \cite{qanli}. The repository of software and converted datasets is available at http://toAppear.

\subsection{QA Pairs to OpenIE Tuples}

We started by examining SQuAD and noticing that each answer, $A$, corresponds to either the subject, relation, or object in an implicit extraction. The corresponding question, $Q$, contains the other two parts, i.e. either the (1) subject and relation, (2) subject and object, or (3) relation and object. Which two pieces the question contains depends on the type of question. For example, ``who was... \textit{factoid}'' type questions contain the relation (``was'') and object (the factoid), which means that the answer is the subject. In Figure 1, ``Who was Rollo'' is recognized as a \textit{who was} question and caught by the \textit{whoParse()} parser. Similarly, a question in the form of ``When did \textit{person} do \textit{action}'' expresses a subject and a relation, with the answer containing the object. For example, ``When did Einstein emigrate to the US`` and answer \textit{1933}, would convert to (Einstein, when did emigrate to the US, 1933). In cases like these the relation might not be grammatically ideal, but nevertheless captures the meaning of the input sentence.

In order to identify generic patterns, we build our parse-based tool on top of a dependency parser \cite{parsingIE}. It uses fifteen rules, with the proper rule being identified and run based on the question type. The rule then uses its pre-specified pattern to parse the input QA pair and output a tuple. These fifteen rules are certainly not exhaustive, but cover around eighty percent of the inputs.  The tool ignores questions greater than 60 characters and complex questions it cannot parse, leaving a dataset smaller than the original (see Table 1).

Each rule is on average forty lines of code that traverses a dependency parse tree according to its pre-specified pattern, extracting the matching spans at each step. A master function \textit{parse()} determines which rule to apply based on the question type which is categorized by nsubj presence, and the type of question (who/what/etc.). Most questions contain an nsubj which makes the parse task easier, as this will also be the subject of the tuple. We allow the master \textit{parse()} method try multiple rules. It first tries very specific rules (e.g. a parser for \textit{how} questions where no subject is identified), then falls down to more generic rules. If no output is returned after all the methods are tried we throw the QA pair out. Otherwise, we find the appropriate sentence in the passage based on the index.

\begin{figure}
    \centering
    \label{fig:norman}
    
\includegraphics[width=\columnwidth]{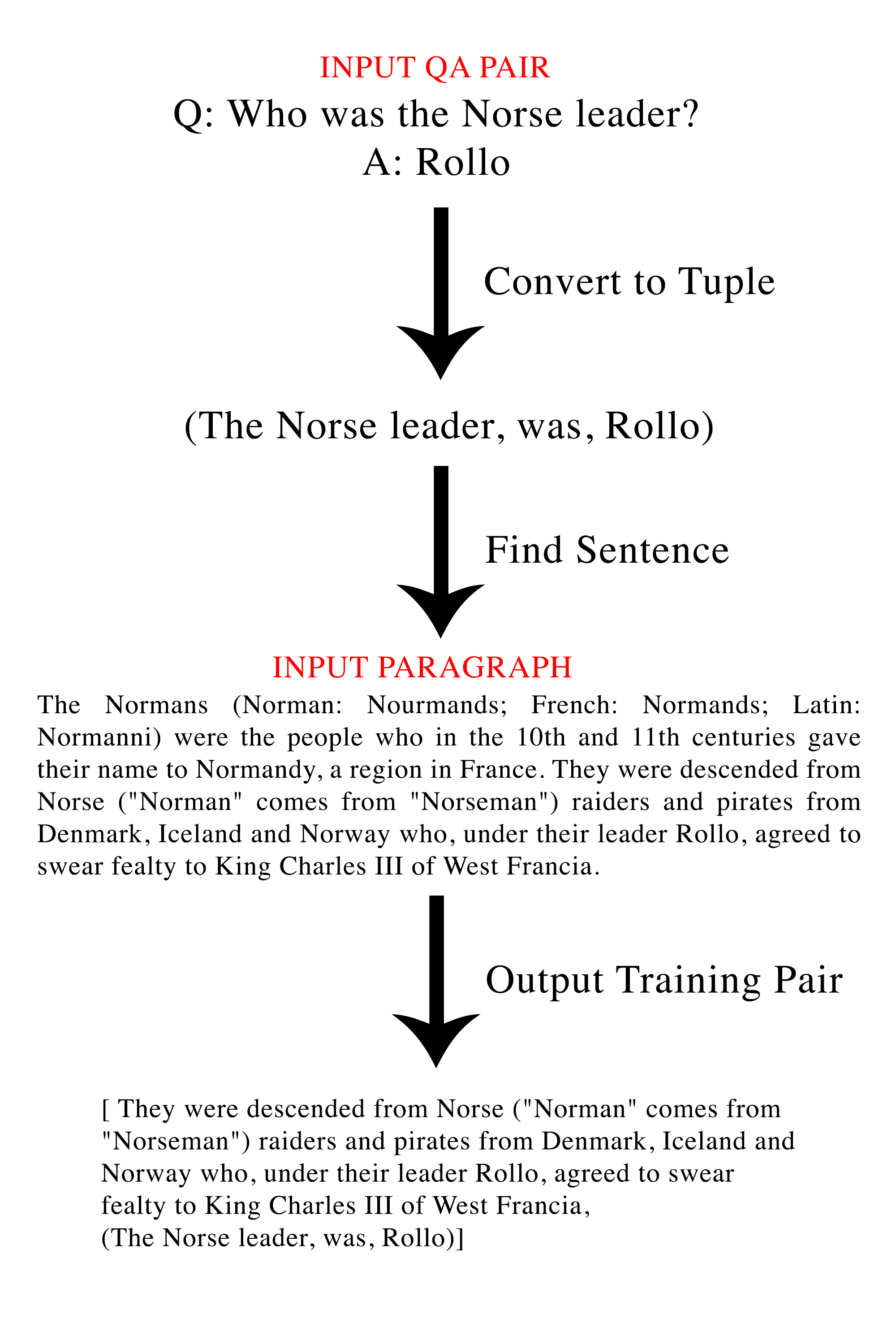}

\caption{Tuple conversion and alignment process flow.}
\end{figure}

\subsection{Sentence Alignment}

Following QA to tuple conversion, the tuple must be aligned with a sentence in the input passage. We segment the passage into sentences using periods as delimiters. The sentence containing the answer is taken as the input sentence for the tuple. Outputted sentences predominantly align with their tuple, but some exhibit partial misalignment in the case of some multi-sentence reasoning questions. 13.6\% of questions require multi-sentence reasoning, so this is an upper bound on the number of partially misaligned tuples/sentences \cite{Squad}. While there may be heuristics that can be used to check alignment, we didn't find a significant number of these misalignments and so left them in the corpus. Figure 1 demonstrates the conversion process.

\subsection{Tuple Examination}

Examining a random subset of one hundred generated tuples in the combined dataset we find 12 noun-mediated, 33 sentence-level inference, 11 long sentence, 7 nominzalization, 0 noisy informal, 3 pp-attachment, 24 explicit, and 10 partially misaligned. With 66\% implicit relations, this dataset shows promise in improving OpenIE's recall on implicit relations.

\section{Our model}

Our implicit OpenIE extractor is implemented as a sequence to sequence model with attention \cite{Bahdanau}. We use a 2-Layer LSTM Encoder/Decoder with 500 parameters, general attention, SGD optimizer with adaptive learning rate, and 0.33 dropout \cite{lstm}. The training objective is to maximize the likelihood of the output tuple given the input sentence. In the case of a sentence having multiple  extractions, it appears in the dataset once for each output tuple. At test time, beam search is used for decoding to produce the top-10 outputs and an associated log likelihood value for each tuple (used to generate the precision-recall curves in Section 7).  

\begin{table}[]
\resizebox{\columnwidth}{!}{%
\begin{tabular}{@{}llll@{}}
\toprule
\textbf{Source Data} & \textbf{Sentences} & \textbf{Train Tuples} & \textbf{Validation Tuples} \\ \midrule
NewsQA & 50880 & 56646 & \multicolumn{1}{c}{-} \\
SQuAD & 38773 & 51949 & \multicolumn{1}{c}{-} \\
Total & 89653 & 107595 & 1000 \\ \bottomrule
\end{tabular}%
}
\caption{\label{font-table} Dataset statistics. }
\end{table}

\section{Evaluation}

We make use of the evaluation tool developed by Stanovsky and Dagan \shortcite{benchmark} to test the precision and recall of our model against previous methods. We make two changes to the tool as described below.

\subsection{Creating a Gold Dataset}

The test corpus contained no implicit data, so we re-annotate 300 tuples from the CoNLL-2009 English training data to use as gold data. Both authors worked on different sentence sets then pruned the other set to ensure only implicit relations remained. We note that this is a different dataset than our training data so should be a good test of generalizability; the training data consists of Wikipedia and news articles, while the test data resembles corporate press release headlines.

\subsection{Matching function for implicit tuples}

We implement a new matching function (i.e. the function that decides if a generated tuple matches a gold tuple). The included matching functions used BoW overlap or BLEU, which aren't appropriate for implicit relations; our goal is to assess whether the meaning of the predicted tuple matches the gold, not the only tokens. For example, the if the gold relation is ``is employed by'' we want to accept ``works for''. Thus, we instead compute the cosine similarity of the subject, relation, and object embeddings to our gold tuple. All three must be above a threshold to evaluate as a match. The sequence embeddings are computed by taking the average of the GloVe embeddings of each word (i.e. BoW embedding) \cite{glove}.

\begin{figure}
    \centering
    \includegraphics[width=\columnwidth]{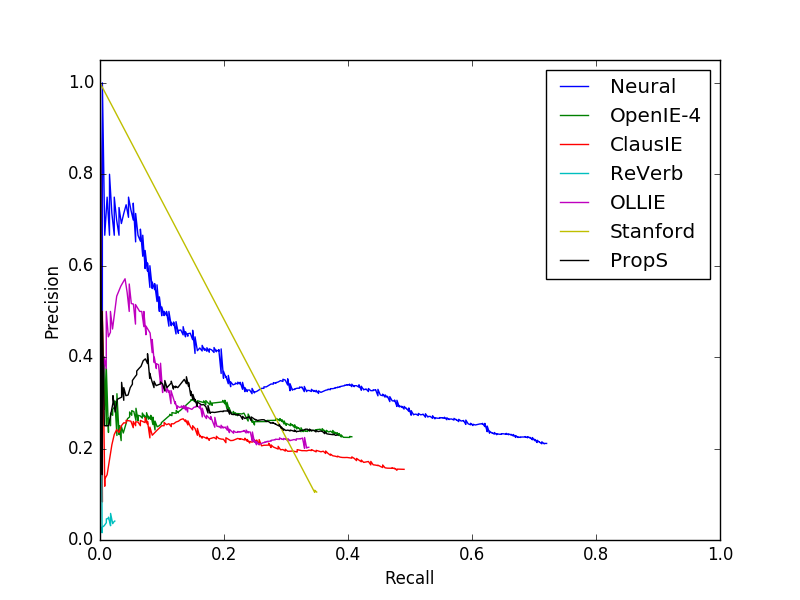}
    \caption{PR curve on our \textbf{implicit} tuples dataset.}
    \label{fig:implicitPR}
\end{figure}

\begin{figure}
    \centering
    \includegraphics[width=\columnwidth]{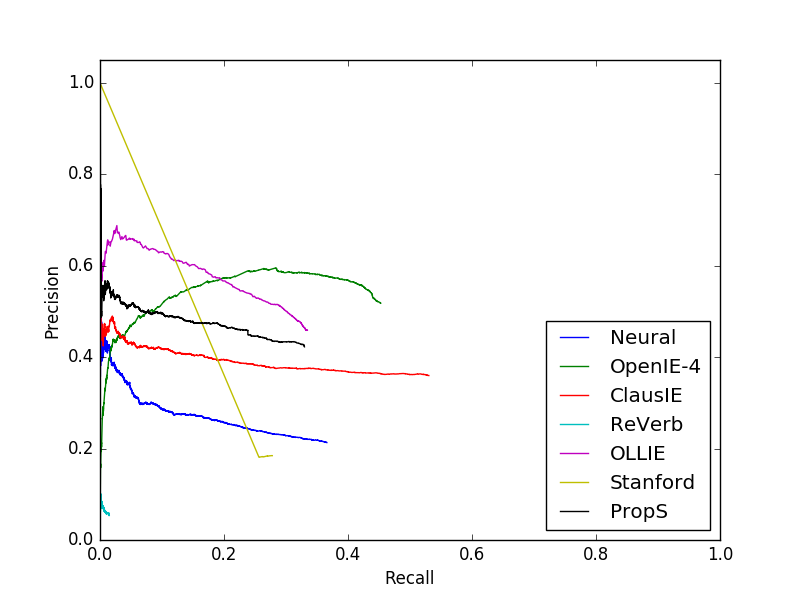}
    \caption{PR curve on the \textbf{explicit} tuples dataset.}
    \label{fig:explicitPR}
\end{figure}

\section{Results}

The results on our implicit corpus are shown in Figure 2 (our method in blue). For continuity with prior work, we also compare our model on the origional corpus but using our new matching function in Figure 3.

Our model outperforms at every point in the implicit-tuples PR curve, accomplishing our  goal of increasing recall on implicit relations. Our system performs poorly on explicit tuples, as we would expect considering our training data. We tried creating a multi-task model, but found the model either learned to produce implit or explicit tuples. Creating a multi-task network would be ideal, though it is sufficient for production systems to use both systems in tandem.

\section{Conclusion}

We created a large training corpus for implicit OpenIE extractors based on SQuAD and NewsQA, trained a baseline on this dataset, and presented promising results on implicit extraction. We see this as part of a larger body of work in text-representation schemes which aim to represent meaning in a more structured form than free text. Implicit information extraction goes further than traditional OpenIE to elicit relations not contained in the original free text. This allows maximally-shortened tuples where common sense relations are made explicit. Our model should improve further as more QA datasets are released and converted to OpenIE data using our conversion tool.


\bibliography{acl2019}

\begin{thebibliography}{24}
\expandafter\ifx\csname natexlab\endcsname\relax\def\natexlab#1{#1}\fi

\bibitem[{Angeli et~al.(2015)Angeli, Premkumar, and Manning}]{stanford}
Gabor Angeli, Melvin Jose~Johnson Premkumar, and Christopher~D. Manning. 2015.
\newblock Leveraging linguistic structure for open domain information
  extraction.
\newblock In \emph{ACL}.

\bibitem[{Bahdanau et~al.(2014)Bahdanau, Cho, and Bengio}]{Bahdanau}
Dzmitry Bahdanau, Kyunghyun Cho, and Yoshua Bengio. 2014.
\newblock Neural machine translation by jointly learning to align and
  translate.
\newblock \emph{CoRR}, abs/1409.0473.

\bibitem[{Banko et~al.(2007)Banko, Cafarella, Soderland, Broadhead, and
  Etzioni}]{origionalOpenIE}
Michele Banko, Michael~J. Cafarella, Stephen Soderland, Matthew~G Broadhead,
  and Oren Etzioni. 2007.
\newblock Open information extraction from the web.
\newblock In \emph{IJCAI}.

\bibitem[{Berant et~al.(2011)Berant, Dagan, and Goldberger}]{entailment}
Jonathan Berant, Ido Dagan, and Jacob Goldberger. 2011.
\newblock Global learning of typed entailment rules.
\newblock In \emph{ACL}.

\bibitem[{Corro and Gemulla(2013)}]{clausie}
Luciano~Del Corro and Rainer Gemulla. 2013.
\newblock Clausie: clause-based open information extraction.
\newblock In \emph{WWW}.

\bibitem[{Cui et~al.(2018)Cui, Wei, and Zhou}]{NeuralOpenIE}
Lei Cui, Furu Wei, and Ming Zhou. 2018.
\newblock Neural open information extraction.
\newblock In \emph{ACL}.

\bibitem[{Demszky et~al.(2018)Demszky, Guu, and Liang}]{qanli}
Dorottya Demszky, Kelvin Guu, and Percy Liang. 2018.
\newblock Transforming question answering datasets into natural language
  inference datasets.
\newblock \emph{CoRR}, abs/1809.02922.

\bibitem[{Fader et~al.(2011)Fader, Soderland, and Etzioni}]{ReVerb2011}
Anthony Fader, Stephen Soderland, and Oren Etzioni. 2011.
\newblock Identifying relations for open information extraction.
\newblock In \emph{Proceedings of the Conference of Empirical Methods in
  Natural Language Processing ({EMNLP} '11)}, Edinburgh, Scotland, UK.

\bibitem[{Hochreiter and Schmidhuber(1997)}]{lstm}
Sepp Hochreiter and J{\"u}rgen Schmidhuber. 1997.
\newblock Long short-term memory.
\newblock \emph{Neural Computation}, 9:1735--1780.

\bibitem[{Honnibal and Johnson(2015)}]{parsingIE}
Matthew Honnibal and Mark Johnson. 2015.
\newblock An improved non-monotonic transition system for dependency parsing.
\newblock In \emph{EMNLP}.

\bibitem[{Klein et~al.(2017)Klein, Kim, Deng, Crego, Senellart, and
  Rush}]{OpenNMT}
Guillaume Klein, Yoon Kim, Yuntian Deng, Josep~Maria Crego, Jean Senellart, and
  Alexander~M. Rush. 2017.
\newblock Opennmt: Open-source toolkit for neural machine translation.
\newblock In \emph{ACL}.

\bibitem[{Mausam et~al.(2012)Mausam, Schmitz, Bart, Soderland, and
  Etzioni}]{ollie}
Mausam, Michael Schmitz, Robert Bart, Stephen Soderland, and Oren Etzioni.
  2012.
\newblock Open language learning for information extraction.
\newblock In \emph{Proceedings of Conference on Empirical Methods in Natural
  Language Processing and Computational Natural Language Learning
  (EMNLP-CONLL)}.

\bibitem[{Niklaus et~al.(2018)Niklaus, Cetto, Freitas, and Handschuh}]{Survey}
Christina Niklaus, Matthias Cetto, Andr{\'e} Freitas, and Siegfried Handschuh.
  2018.
\newblock A survey on open information extraction.
\newblock In \emph{COLING}.

\bibitem[{Pal and Mausam(2016)}]{Demonyms}
Harinder Pal and Mausam. 2016.
\newblock Demonyms and compound relational nouns in nominal open ie.
\newblock In \emph{AKBC@NAACL-HLT}.

\bibitem[{Pennington et~al.(2014)Pennington, Socher, and Manning}]{glove}
Jeffrey Pennington, Richard Socher, and Christopher~D. Manning. 2014.
\newblock Glove: Global vectors for word representation.
\newblock In \emph{EMNLP}.

\bibitem[{Rajpurkar et~al.(2016)Rajpurkar, Zhang, Lopyrev, and Liang}]{Squad}
Pranav Rajpurkar, Jian Zhang, Konstantin Lopyrev, and Percy Liang. 2016.
\newblock Squad: 100, 000+ questions for machine comprehension of text.
\newblock In \emph{EMNLP}.

\bibitem[{Saha and Mausam(2018)}]{conjunctiveIE}
Swarnadeep Saha and Mausam. 2018.
\newblock Open information extraction from conjunctive sentences.
\newblock In \emph{COLING}.

\bibitem[{Saha et~al.(2017)Saha, Pal, and Mausam}]{Bootstrapping}
Swarnadeep Saha, Harinder Pal, and Mausam. 2017.
\newblock Bootstrapping for numerical open ie.
\newblock In \emph{ACL}.

\bibitem[{Stanovsky and Dagan(2016)}]{benchmark}
Gabriel Stanovsky and Ido Dagan. 2016.
\newblock Creating a large benchmark for open information extraction.
\newblock In \emph{EMNLP}.

\bibitem[{Stanovsky et~al.(2015)Stanovsky, Dagan, and Mausam}]{IEintermediate}
Gabriel Stanovsky, Ido Dagan, and Mausam. 2015.
\newblock Open ie as an intermediate structure for semantic tasks.
\newblock In \emph{ACL}.

\bibitem[{Stanovsky et~al.(2016)Stanovsky, Ficler, Dagan, and Goldberg}]{propS}
Gabriel Stanovsky, Jessica Ficler, Ido Dagan, and Yoav Goldberg. 2016.
\newblock Getting more out of syntax with props.
\newblock \emph{CoRR}, abs/1603.01648.

\bibitem[{Stanovsky et~al.(2018)Stanovsky, Michael, Zettlemoyer, and
  Dagan}]{SupervisedOIE}
Gabriel Stanovsky, Julian Michael, Luke~S. Zettlemoyer, and Ido Dagan. 2018.
\newblock Supervised open information extraction.
\newblock In \emph{NAACL-HLT}.

\bibitem[{Trischler et~al.(2017)Trischler, Wang, Yuan, Harris, Sordoni,
  Bachman, and Suleman}]{newsQA}
Adam Trischler, Tong Wang, Xingdi Yuan, Justin Harris, Alessandro Sordoni,
  Philip Bachman, and Kaheer Suleman. 2017.
\newblock Newsqa: A machine comprehension dataset.
\newblock In \emph{Rep4NLP@ACL}.

\bibitem[{Wities et~al.(2017)Wities, Shwartz, Stanovsky, Adler, Shapira,
  Upadhyay, Roth, Camara, Gurevych, and Dagan}]{okr}
Rachel Wities, Vered Shwartz, Gabriel Stanovsky, Meni Adler, Ori Shapira, Shyam
  Upadhyay, Dan Roth, Eugenio~Martinez Camara, Iryna Gurevych, and Ido Dagan.
  2017.
\newblock A consolidated open knowledge representation for multiple texts.
\newblock In \emph{LSDSem}.

\end{thebibliography}
\bibliographystyle{acl_natbib}

\end{document}